\documentclass[aps,prl,twocolumn,showpacs]{revtex4-1}
\usepackage{amsmath}
\usepackage{amssymb}
\usepackage{graphicx}
\usepackage{epstopdf}
\DeclareMathOperator{\sgn}{sgn}

\newcommand{\ud}{{\uparrow \downarrow}}

\begin{document}
\title{Topological spin transport by Brownian diffusion of domain walls}

\author{Se Kwon Kim}
\affiliation{
	Department of Physics and Astronomy,
	University of California,
	Los Angeles, California 90095, USA
}

\author{So Takei}
\affiliation{
	Department of Physics and Astronomy,
	University of California,
	Los Angeles, California 90095, USA
}

\author{Yaroslav Tserkovnyak}
\affiliation{
	Department of Physics and Astronomy,
	University of California,
	Los Angeles, California 90095, USA
}

\date{\today}

\begin{abstract}
We propose thermally-populated domain walls (DWs) in an easy-plane ferromagnetic insulator as robust spin carriers between two metals. The chirality of a DW, which serves as a topological charge, couples to the metal spin accumulation via spin-transfer torque and results in the chirality-dependent thermal nucleation rates of DWs at the interface. After overpopulated DWs of a particular (net) chirality diffuse and leave the ferromagnet at the other interface, they reemit the spin current by spin pumping. The conservation of the topological charge supports an algebraic decay of spin transport as the length of the ferromagnet increases; this is analogous to the decaying behavior of superfluid spin transport but contrasts with the exponential decay of magnon spin transport. We envision that similar spin transport with algebraic decay may be implemented in materials with exotic spin phases by exploiting topological characteristics and the associated conserved quantities of their excitations.
\end{abstract}


\maketitle

\emph{Introduction.}|Spintronics, or spin-transport electronics, exploits spin degrees of freedom in condensed matter systems to improve information processing technology that is traditionally based on electric charge \cite{WolfScience2001, *ZuticRMP2004}. Conducting materials have been used to transport spin by polarizing itinerant electrons, which is associated with undesired energy dissipation due to the electronic continuum. Magnetic insulators, which are immune to Joule heating, provide alternative platforms to seek an efficient spin transport channel. Superfluid spin transport \cite{SoninJETP1978, *SoninAP2010, KonigPRL2002, *ChenPRB2014, ChenPRB2014-2, *ChenPRL2015, TakeiPRL2014, *TakeiPRB2014} has been proposed for long-ranged spin transmission in magnetic insulators with easy-plane anisotropy. The spin superfluidity, however, can be destroyed by U(1)-symmetry-breaking anisotropy within the easy plane.

Topological solitons in magnetic materials are nonlinear excitations that are protected by their nontrivial topology \cite{KosevichPR1990, *BraunAP2012}. A domain wall (DW) in an easy-axis magnet is a prototypical topological soliton, which can store and deliver information as demonstrated in the racetrack memory \cite{ParkinScience2008}. DWs can be driven by various means, e.g., an external magnetic field \cite{SchryerJAP1974}, an electric current (in conducting systems) \cite{BergerJAP1978}, or heat flux \cite{BergerJAP1985, *JenJAP1985, *JenJAP1985-2, HatamiPRL2007, SaslowPRB2007, HalsSSC2010, HinzkePRL2011, YanPRL2011, JiangPRL2013, *ChicoPRB2014}. At a finite temperature, DWs with damped dynamics undergo Brownian motion due to a random force dictated by the fluctuation-dissipation theorem \cite{LL5, NeelCRASP1949, *NeelAG1949, *BrownPR1963, *KuboPTPS1970, OgataJPSJ1986, AlessandroJAP1990, *DuinePRL2007, *SchuttePRB2014, IvanovJPCM1993}; under a temperature gradient, Brownian motion leads to a diffusive transport (thermophoresis) of DWs \cite{KimPRB2015}.

\begin{figure}
\includegraphics[width=0.95\columnwidth]{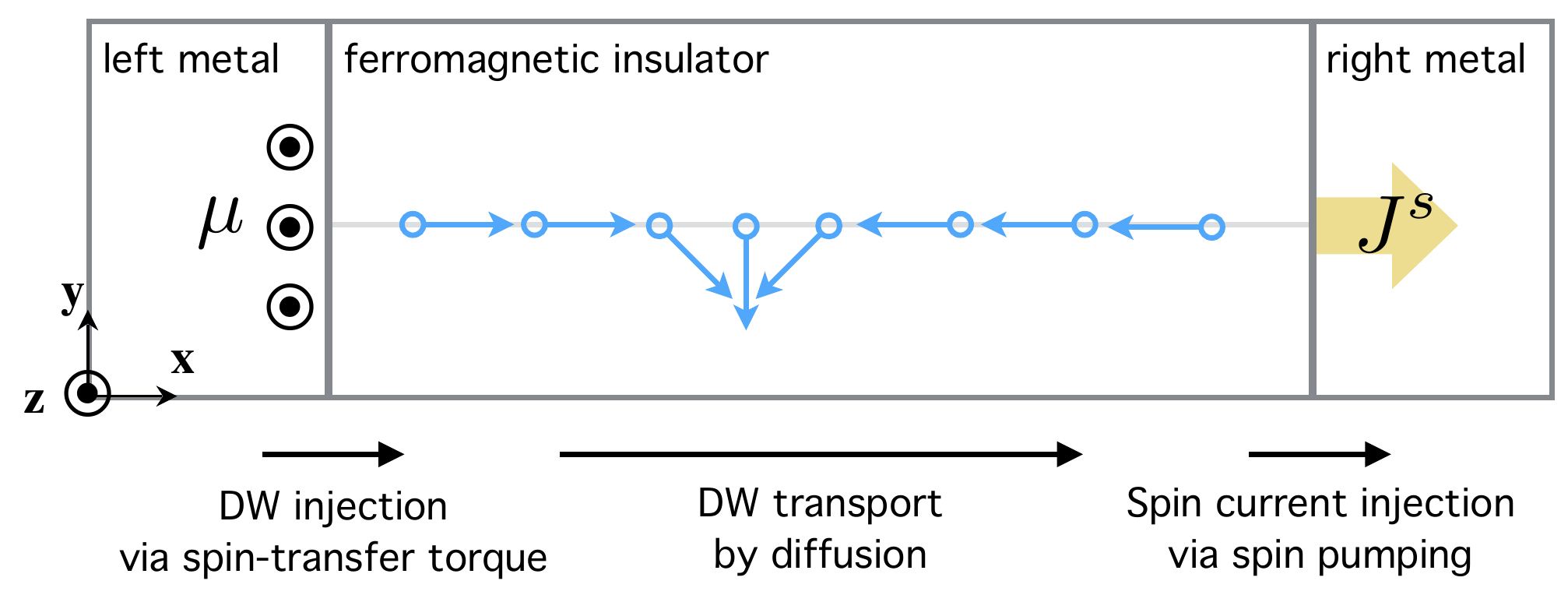}
\caption{(color online). An easy-$xy$-plane ferromagnetic insulator with an additional easy-axis anisotropy in the $x$ direction is sandwiched between two metals. The spin-transfer torque caused by the out-of-equilibrium spin accumulation $\mu$ in the positive $z$ direction prefers injection of DWs with the clockwise-rotating magnetization. The annihilation of these DWs generates the spin current into the right metal via spin pumping. In the diffusive limit of DW motion, the spin current decays algebraically as the ferromagnet's length increases.}
\label{fig:fig1}
\end{figure}

In this Rapid Communication, we show that superfluid-like spin transport can be achieved by utilizing thermally-populated DWs in an easy-plane ferromagnetic insulator with an additional easy-axis anisotropy within the easy plane. Long thin ferromagnetic strips, for example, are naturally endowed with such anisotropies due to magnetostatic interactions \cite{OsbornPR1945, SchryerJAP1974}. See Fig.~\ref{fig:fig1} for illustration. A DW is characterized by its chirality $q = \pm 1$, associated with the sense of circulation of the magnetization within the easy plane \cite{KosevichPR1990}. The chirality of a DW is protected in the $XY$ ferromagnet by topology, and we thus refer to it as the topological ``charge." Suppose the ferromagnet is driven out of equilibrium by the spin accumulation in the positive $z$ direction in the left metal. The induced spin-transfer torque nucleates DWs with the clockwise-rotating magnetization. When these DWs leave the ferromagnet toward the right metal, the magnetization at the interface rotates counterclockwise, which, in turn, generates the spin current into the metal via spin pumping. In the diffusive regime of DW motion, the spin current transported by DWs decays algebraically as in superfluid spin transport \cite{TakeiPRL2014} owing to the conservation of the topological charge. This topological spin transport can be inferred by measuring the drag coefficient in a magnetoelectric circuit that was proposed in Ref.~\cite{TakeiPRL2014} for detecting superfluid spin transport.

\begin{figure}
\includegraphics[width=\columnwidth]{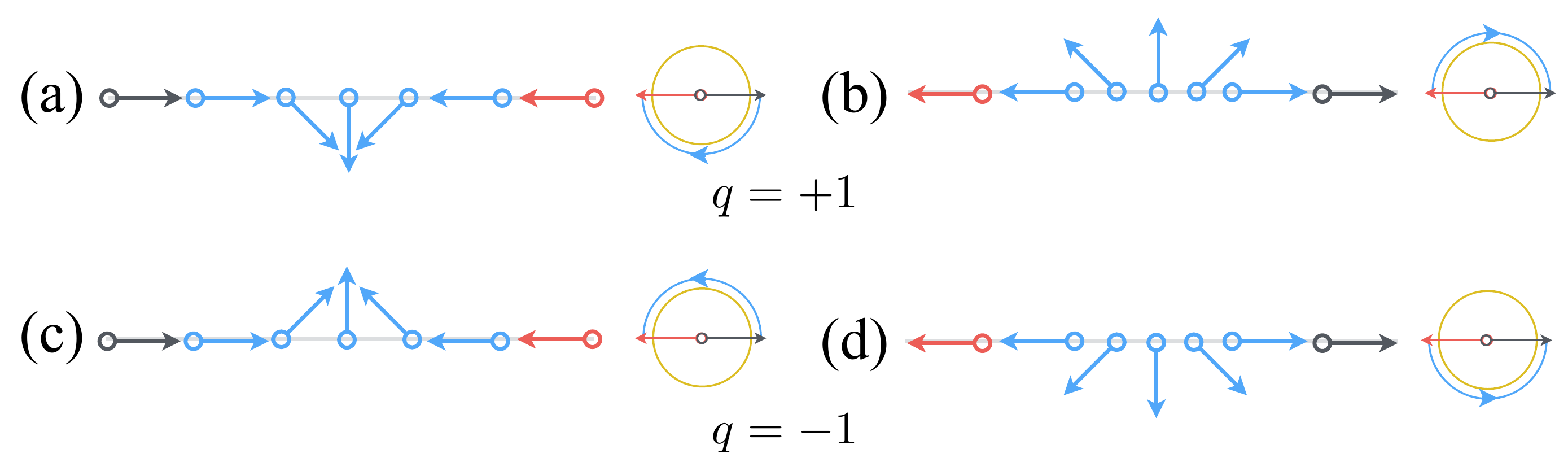}
\caption{(color online). (a), (b) The DWs with the topological charge $q = 1$. (c), (d) The walls with $q = -1$.}
\label{fig:fig2}
\end{figure}

\emph{Main results.}|The model system consists of a quasi one-dimensional easy-$xy$-plane ferromagnetic insulator with an additional easy-$x$-axis anisotropy attached on both sides by nonmagnetic metals. In equilibrium, the anisotropy lays the local spin density $\mathbf{s} \equiv s \mathbf{n}$ in the $xy$ plane, which allows us to parametrize its direction as $\mathbf{n} = (\cos \phi, \sin \phi, 0)$. A DW is a topologically stable equilibrium texture that interpolates the two uniform ground states, $\phi \equiv 0$ or $\pi$. Its associated winding is characterized by the topological charge:
\begin{equation}
q \equiv - \frac{1}{\pi} \int \mathrm{d}x \, \partial_x \phi \, ,
\label{eq:q}
\end{equation}
where the integral is over the DW along the longitudinal $x$ axis of the ferromagnet. Figure~\ref{fig:fig2} illustrates four possible DW types.

\begin{figure}
\includegraphics[width=\columnwidth]{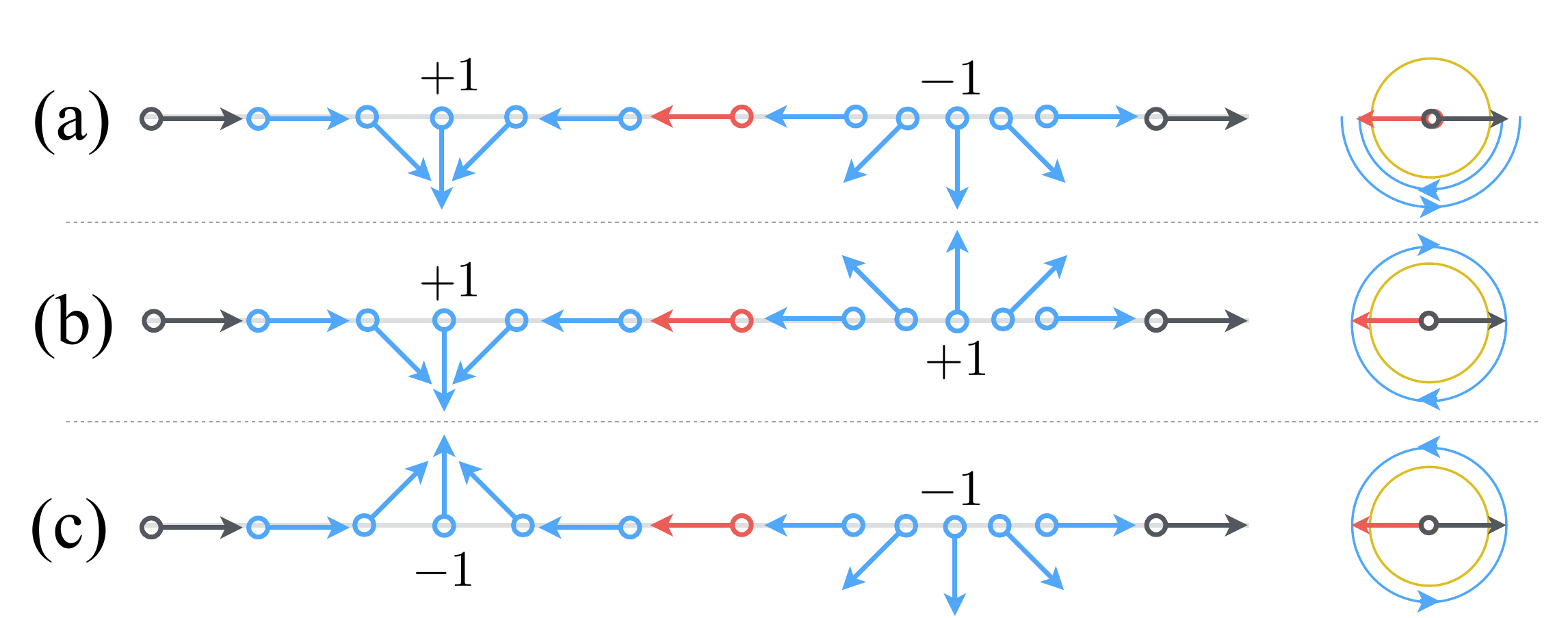}
\caption{(color online). Schematic for the conservation of the total topological charge. (a) A pair of DWs with opposite charges, so that the direction of the magnetization does not wind the circle as shown in the right. The magnetization texture is, therefore, topologically trivial and can be created or destroyed spontaneously. (b), (c) A pair of DWs with the same charge. The direction of the magnetization winds around the circle once, which makes the textured configuration topologically stable from thermal annihilation. The total topological charge, i.e., the net winding number, is conserved during interactions between DWs.}
\label{fig:fig3}
\end{figure}

A finite temperature causes spontaneous nucleation and annihilation of DWs. In the bulk, DWs are created and destroyed always in pairs with opposite charges as shown in Fig.~\ref{fig:fig3}(a) \footnote{At high temperatures, there is also a possibility for thermally-activated phase slips. These yield DWs of the same charge, thus invalidating conservation of the topological charge in the bulk \cite{KimarXiv2015}. We are disregarding phase slips by focusing on an intermediate-temperature regime: $T \ll S \sqrt{A K}$, energy barrier for phase slips. The temperature, however, should be comparable to the DW energy, $T \lesssim S \sqrt{A \kappa}$, to avoid exponential suppression of nucleation of DWs. The condition for the consistency between these two criteria is $\kappa \ll K$. This is naturally supported by the shape anisotropies of an infinitely long thin magnetic strip, $\kappa / K \simeq t / w \ll 1$ with $t$ the thickness and $w$ the width of the strip \cite{OsbornPR1945}.}. The topological charge density, $\rho \equiv \rho^+ - \rho^-$ is, thus, preserved in the bulk [Fig.~\ref{fig:fig3}(b) and (c)], where $\rho^\pm$ are the linear densities of DWs with $q = \pm 1$, respectively. A topological charge can be injected or ejected through the boundaries of the ferromagnet. In equilibrium, the DW density is charge-independent; $\rho^\pm \rightarrow \rho_0 \propto \exp (- E_0 / T)$, where $E_0$ is the DW energy.

A DW should generally behave as a particle immersed in a viscous medium due to its coupling to, e.g., lattice vibrations \cite{BrownPR1963} or other microscopic degrees of freedom. As such, it must exhibit Brownian motion at a finite temperature due to random forces, whose existence is dictated by the fluctuation-dissipation theorem \cite{LL5}. For a conglomerate of DWs that diffuse by Brownian motion, the dynamics of the topological charge density is described by the Fokker-Planck equation \cite{KimPRB2015}:
\begin{equation}
\partial_t \rho + \partial_x I = 0, \quad I \equiv - D \partial_x \rho \, ,
\label{eq:fp}
\end{equation}
in the absence of an external force, where $I$ is the topological charge current. In equilibrium, the density and the current of the topological charge are zero; $\rho = 0 = I$ according to the reflection symmetry in the $xz$ plane and the time reversal symmetry.

The topological charge density can be injected by perturbing the ferromagnet by the nonequilibrium $z$ axis spin accumulation in the left metal,  $\boldsymbol{\mu} \equiv \mu \hat{\mathbf{z}}$, which we assume positive, $\mu > 0$, for concreteness. The spin-transfer torque caused by the spin accumulation is $\boldsymbol{\tau} = (g_L' + g_L \mathbf{n} \times) (\boldsymbol{\mu} \times \mathbf{n}) / 4 \pi$, where $g_i^\ud \equiv g_i + \imath g_i'$ is the effective complex spin mixing conductance associated with the ferromagnet/metal-$i$ interface \cite{TserkovnyakRMP2005}. The torque does work on the ferromagnet favoring the nucleation of DWs with the positive charge: $W^q = q g_L \mu S / 4$, where $q$ is the charge of the wall and $S$ is the cross-sectional area of the ferromagnet. The resultant nucleation rate of the topological charge is $\Gamma_L \delta W / T$ to linear order in the bias, where $\Gamma_L$ is the equilibrium-nucleation rate of DWs at the left interface and $\delta W \equiv W^+ - W^- = g_L \mu S / 2$ is the difference between the two works.

The injected topological charges diffuse by Brownian motion and can leave the ferromagnet through the right boundary. The conservation of the topological charge leads to the steady-state current (as derived below):
\begin{equation}
I = \frac{g_L \mu}{R_L + R_R + R_B} \, , \label{eq:Itc}
\end{equation}
where 
\begin{equation}
R_L \equiv \frac{2 T}{\Gamma_L S}, \quad R_R \equiv \frac{2 T}{\Gamma_R S}, \quad R_B \equiv \frac{2 T L}{\rho_0 D S} \, ,
\end{equation}
and $L$ is the length of the ferromagnet. We may interpret the topological charge current $I$ as the applied ``voltage" $g_L \mu$ (with units of $\text{J}/\text{m}^2$) divided by the total ``resistance" $R_L + R_R + R_B$ (with units of $\text{J} \cdot \text{s} / \text{m}^2$) of the series circuit, which is made of the interface resistances, $R_L$ and $R_R$, and the bulk resistance $R_B$. Note that the bulk resistance $R_B$ is proportional to the ratio of the length to the cross-sectional area, $L / S$, which is analogous to the electrical resistance.

The dynamics of the local spin density at the boundaries injects spin current into the metals via spin pumping, which is the Onsager reciprocal effect \cite{LL5} to spin-transfer torque. The spin current density associated with spin pumping at the right interface is $\mathbf{J}^s_R = \hbar (g_R' + g_R \mathbf{n} \times) \dot{\mathbf{n}} / 4 \pi$. The annihilation of the topological charge pumps spin current polarized in the $z$ direction to the right metal:
\begin{equation}
J_R^s = \frac{\hbar g_R}{4} I = \frac{\hbar g_R g_L \mu}{4 (R_L + R_R + R_B)} \, . \label{eq:main}
\end{equation}
This is a central result of our work. Note that the spin current decays algebraically as a function of the ferromagnet's length $L$, which is similar to superfluid spin transport in an easy-plane ferromagnet \cite{TakeiPRL2014}, but contrasts with the exponential decay of the spin transport by thermal magnons \cite{ZhangPRL2012}. The formalism that we have developed is general enough to be readily extended to other easy-plane magnets, e.g., the case of an antiferromagnet with an additional easy-axis anisotropy within the easy-plane is closely analogous \cite{TakeiPRL2014}.

\emph{Brownian motion.}|Let us provide an explicit model for Brownian motion of DWs following Ref.~\cite{KimPRB2015}. We assume the following free energy for the ferromagnet: $U[\mathbf{n}] = \int dV (A |\partial_x \mathbf{n}|^2 + K n_z^2 - \kappa n_x^2) / 2$, where $A$ represents the exchange stiffness, and the positive coefficients $\kappa$ and $K$ parameterize the anisotropy magnitudes. In equilibrium, the local spin density $\mathbf{s} = s \mathbf{n}$ lies in the $xy$ plane, which can be parametrized by its azimuthal angle $\phi$. A static DW solution centered at $X$ is given by
\begin{equation}
\cos [\phi_0 (x - X)] = \pm \tanh[(x - X) / \Delta] \, ,
\label{eq:dw}
\end{equation}
with the chirality of the DW given by $\sgn( \cos \phi_0 \cdot \sin \phi_0 )|_{x > X}$, the energy $E_0 = 2 S \sqrt{A \kappa}$, and the width $\Delta = \sqrt{A/\kappa}$ \cite{SchryerJAP1974}. We assume here and hereafter that the ambient temperature is much lower than the ordering temperature, $T \ll T_c$, for which thermally-induced changes of DW properties can be ignored. Figure~\ref{fig:fig2} depicts possible types of DWs. 

The dynamics of $\mathbf{n}$ at a finite temperature is described by the stochastic Landau-Lifshitz-Gilbert (LLG) equation,
\begin{equation}
s (1 +  \alpha \mathbf{n} \times) \dot{\mathbf{n}} = \mathbf{n} \times (\mathbf{h} + \mathbf{h}^\text{th}) \, ,
\end{equation}
where $\mathbf{h} \equiv - \partial U / \partial \mathbf{n}$ is the effective field conjugate to $\mathbf{n}$ and $\mathbf{h}^\text{th}$ is the stochastic Langevin field \cite{BrownPR1963}. The fluctuation-dissipation theorem relates the Gilbert damping constant to the correlator of the Langevin fields; $\langle h^\text{th}_i (\mathbf{r}, t) h^\text{th}_j (\mathbf{r}', t') \rangle = 2 \alpha s T \delta(\mathbf{r} - \mathbf{r}') \delta(t - t')$. The Langevin equation for the overdamped dynamics of $X$ can be obtained from the stochastic LLG equation by the collective coordinate approach \cite{TretiakovPRL2008}:
\begin{equation}
\dot{X} = \eta^{-1} F + v^\text{th} \, ,
\end{equation}
where $\eta \equiv \alpha s \int dV (\partial_x \phi_0)^2 = 2 \alpha s S / \Delta$ is the viscous coefficient, $F \equiv - \partial U / \partial X$ is the conservative force conjugate to $X$, and $v^\text{th} \equiv - \eta^{-1} \int dV (\mathbf{h}^\text{th} \cdot \partial_x \mathbf{n})$ is the stochastic velocity \cite{Pathria1996}. The diffusion coefficient $D$ in the correlator of the stochastic velocity, $\langle v^\text{th} (X, t) v^\text{th} (X', t') \rangle = 2 D \delta(X - X') \delta(t - t')$, is related to the viscous coefficient $\eta$ according to the Einstein-Smoluchowski relation: $D = T / \eta = \Delta T / 2 \alpha s S$ (we set $k_B = 1$).

\emph{Nucleation and annihilation.}|In the bulk of the ferromagnet, DWs are nucleated and annihilated always in pairs with opposite charges [Fig.~\ref{fig:fig3}(a)], which preserves the topological charge density \cite{Note1}. The source and the drain of the topological charge, therefore, can be located only at the boundaries of the ferromagnet. Following the reaction-rate theory \cite{HanggiRMP1990}, the injection rate of DWs through each boundary is given by
\begin{equation}
I^\pm = \Gamma^\pm (T, \mu) - \gamma^\pm (T) \rho^\pm \, , 
\end{equation}
where $\Gamma^q (T, \mu)$ is the nucleation rate, $\gamma^q (T)$ is the annihilation rate per unit density, and $\rho^q$ is the density of $q$-charged DWs.

The annihilation rate per unit density $\gamma^q (T)$ is the characteristic velocity parametrizing the escape of DWs, which, we expect, does not depend on the charge of DWs: $\gamma^q (T) = \gamma (T)$. Interpreting the width $\Delta$ as the mean free path of DWs yields the mean thermal speed $\gamma(T) \sim D(T) / \Delta$.

The nucleation rate of DWs at each interface is $\Gamma^q (T, \mu) = \nu(T) \exp[- E^q (\mu) / T]$, where $E^q (\mu)$ is the energy barrier for the entering of a $q$-charged DW and $\nu (T)$ is the characteristic frequency that depends on details of the system \footnote{$\nu(T)$ is analogous to the attempt frequency in the N\'eel-Brown theory for magnetization reversal of a monodomain ferromagnetic particle \cite{NeelCRASP1949}.}. We take $\nu(T)$ to be independent of the spin accumulation $\mu$ \footnote{The attempt frequency is determined by the curvature of the free energy with respect to fluctuations of the magnetization at the equilibrium point \cite{HanggiRMP1990}. The spin accumulation acts as an external torque, which couples linearly to the ferromagnet and thus does not change the curvature of the free energy. We expect, therefore, for the spin accumulation to not affect the attempt frequency.}, similarly to the N\'eel-Brown theory for thermal switching of magnetic nanoislands subjected to an electrical current \cite{EltschkaPRL2010}.

\emph{Topological spin transport.}|In the presence of a finite spin accumulation $\boldsymbol{\mu} = \mu \hat{\mathbf{z}}$ in the left metal, the energy barrier necessary for the injection of a $q$-charged DW is given by
\begin{equation}
E^q	= E_0 + S \int dx \, \boldsymbol{\tau} \cdot (\delta \phi \, \hat{\mathbf{z}}) = E_0 - q S g_L \mu / 4
\end{equation}
to linear order in $\mu$, from which we obtain the charge-dependent work $W^q = q S g_L \mu / 4$ done by the spin-transfer torque. 

When the spin accumulation is in the positive $z$ direction, $\mu > 0$, the entering of DWs with the positive charge $q = 1$ is favored over $q=-1$. The nucleation rates are $\Gamma^\pm (T, \mu) = \Gamma (T) (1 + W^\pm / T)$ to linear order in $\mu$, where $\Gamma(T) \equiv \Gamma(T, \mu = 0)$ is the equilibrium nucleation rate. The injection rate of the topological charge through the left interface is given by
\begin{equation}
I_L \equiv I^+_L - I^-_L = \Gamma_L (T) \delta W / T - \gamma_L (T) \rho_L \, .
\end{equation}
In the bulk, the topological charge current is $I = - D \partial_x \rho$ from the Fokker-Planck equation~(\ref{eq:fp}).

At the right interface, in the absence of the nonequilibrium spin accumulation, the nucleation rate of a DW is independent of the charge; the topological charge does not enter the ferromagnet, but only leaves it. The topological charge current is, therefore, given by
\begin{equation}
I_R = \gamma_R (T) \rho_R \, .
\end{equation}

The conservation of the topological charge density, $I = I_L = I_R$, leads to the steady-state solution with the uniform topological charge current,
\begin{equation}
I = \frac{\rho_0}{\gamma_L^{-1} + \gamma_R^{-1} + L/D} \frac{\delta W}{T} \, ,
\end{equation}
which can be recast as Eq.~(\ref{eq:Itc}). $\gamma_L$ and $\gamma_R$ are the average velocity of a topological charge to cross the left and right interface, respectively; $D/L$ is the average velocity of a topological charge traversing the ferromagnet, which can be seen from $D / L = L / \delta t$, where $\delta t$ is the average time for a DW to travel the distance $L$.

The spin current density by spin pumping through the right interface is $\mathbf{J}^s_R = \hbar (g_R' + g_R \mathbf{n} \times) \dot{\mathbf{n}} / 4 \pi$. Its $z$ component is $J^s_R \equiv \hat{\mathbf{z}} \cdot \mathbf{J}^s_R = \hbar g_R \dot{\phi} / 4 \pi$ to linear order in the bias. In the steady state with the current $I$ of the charge density $- \partial_x \phi / \pi$, time evolution of the azimuthal angle $\phi$ is given by
\begin{equation}
\dot{\phi}  / \pi = I \, .
\end{equation}
The resultant spin current is $J^s_R = \hbar g_R I / 4$ in Eq.~(\ref{eq:main}).

\emph{Quantitative estimates.}|For quantitative estimates, let us take following parameters for YIG \cite{BhagatPSS1973, TakeiPRL2014}: the spin angular momentum density $s = 10 \, \hbar$/nm$^3$, the Gilbert damping constant $\alpha = 10^{-4}$, and the stiffness coefficient $A = 5 \times 10^{-12} \, $J/m. Long thin YIG strips with thickness $t = 2$ nm and width $w = 50$ nm are given (by the dipolar energy) the shape anisotropy parametrized by $K = 4 \times 10^4 \, $J/m$^3$ and $\kappa = 2 \times 10^3 \, $J/m$^3$ \footnote{$K = 4 \pi M_s^2 w / (t + w)$ and $\kappa = 4 \pi M_s^2 t / (t + w)$ in Gaussian units, where $M_s$ is the saturation magnetization. The demagnetizing factors are for infinitely long wire with an elliptical cross section characterized by two semi axes $t$ and $w$ \cite{OsbornPR1945}.}. 

The algebraical decaying of topological spin transport manifests clearly when the ferromagnet's length $L$ is much larger than the crossover length $L^* \equiv D / \gamma$, for which the bulk resistance dominates the interface ones, $R_B \gg R_L, R_R$. The annihilation rate is estimated as $\gamma \sim D / \Delta$, which yields the crossover length $L^* \sim \Delta = 60 \text{ nm}$. The Boltzmann factor is $\exp(-E_0 / T) \sim 10^{-2}$ at room temperature $T = 300$ K.

\begin{figure}
\includegraphics[width=\columnwidth]{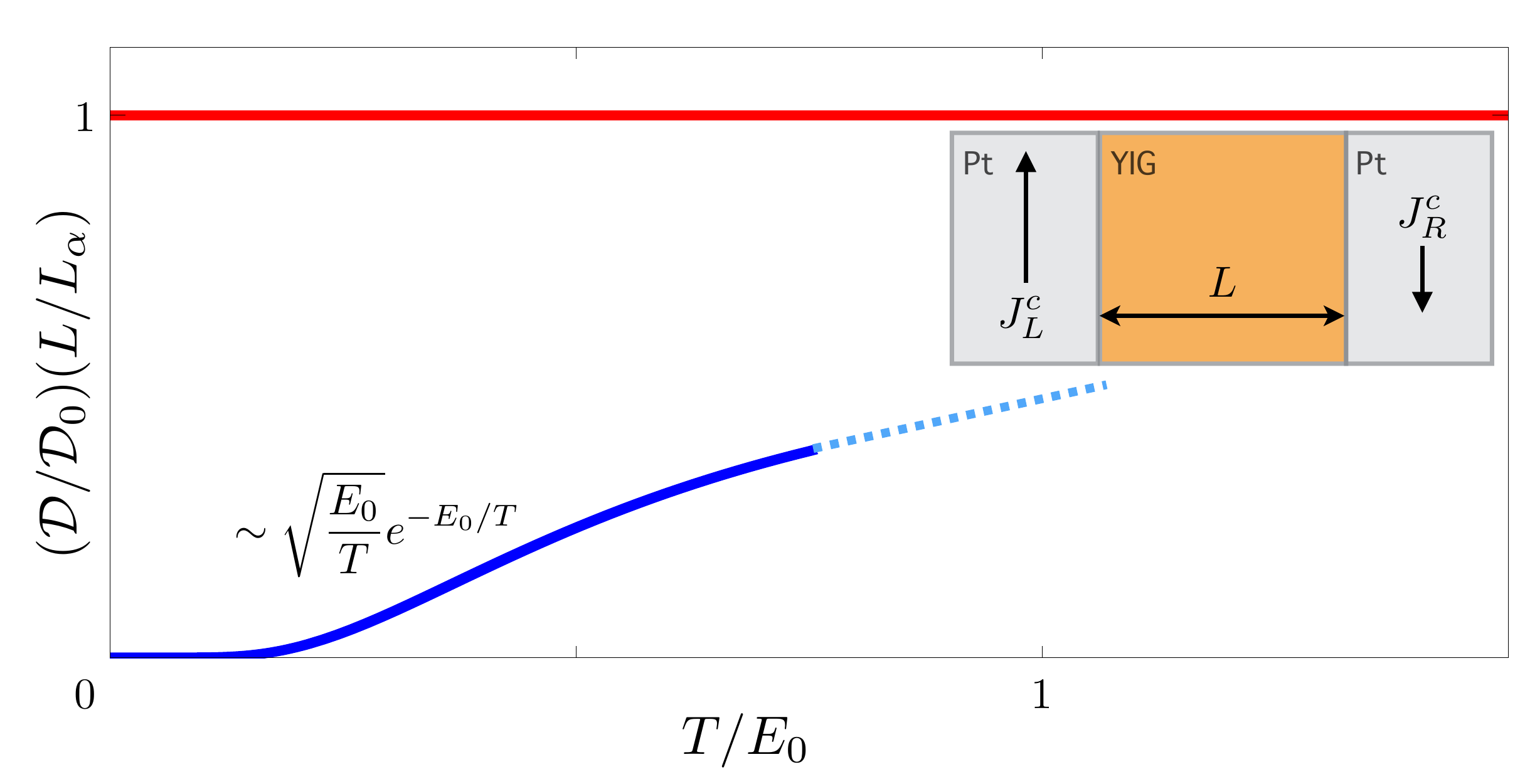}
\caption{(color online). Schematic plot of the drag coefficient $\mathcal{D} = - J^c_R / J^c_L$ normalized by $\mathcal{D}_0 (L_\alpha / L)$ as a function of the ambient temperature $T / E_0$, where $E_0$ is the DW energy. The red straight line is for the absence of anisotropy within the easy plane, $\kappa = 0$. The blue solid curve is for the presence of an easy-axis anisotropy, $\kappa > 0$ at low temperatures $T < 3 E_0 / 4$. The blue dotted line shows speculative extrapolation of the solid curve to higher temperatures $T \sim E_0$.
}
\label{fig:fig4}
\end{figure}

\emph{Discussion.}|The topological spin transport by diffusion of DWs can be experimentally detected in a hybrid structure consisting of a ferromagnetic insulator and two identical strong spin-orbit coupled metals, such as Pt\textbar YIG\textbar Pt (see Fig.~\ref{fig:fig4}), as proposed for superfluid spin transport \cite{TakeiPRL2014}. Given the applied electric-current density $J^c_L$ in the left metal, the spin-current injection into the right metal induces the electric-current density $J^c_R$ via the inverse spin Hall effect, which defines the (negative) drag coefficient, $\mathcal{D} \equiv - J^c_R / J^c_L$. 

Figure~\ref{fig:fig4} schematically depicts the drag coefficient as a function of a temperature in two cases: the presence and absence of an easy-axis anisotropy within the easy plane, $\kappa = 0$ and $\kappa > 0$. We focus on sufficiently long magnetic wires, for which algebraic decaying is prominent: $L \gg L_\alpha$ for $\kappa = 0$ and $L \gg L^*$ for $\kappa > 0$, where $L_\alpha \equiv \hbar g / 2 \pi \alpha s$ ($\sim 1 \text{ $\mu$m}$ for YIG) and $g$ is the real part of the effective mixing conductance of the metal \cite{TakeiPRL2014}. For $\kappa = 0$, superfluid spin transport is sustained by a planar spiraling texture of the magnetization. The drag coefficient is independent of a temperature; $\mathcal{D} = \mathcal{D}_0 (L_\alpha / L)$ with $\mathcal{D}_0 \sim 0.1$ for $1$ nm thick platinum \cite{TakeiPRL2014}. For $\kappa > 0$, superfluid-like spin transport is realized by Brownian diffusion of DWs. Using $D = T \Delta / 2 \alpha s S$, the drag coefficient is $\mathcal{D} (T) = \pi^2 \Delta \rho_0(T) \mathcal{D}_0 (L_\alpha / L)$ for dilute DWs, $T \ll E_0$. The density of DWs is given by $\rho_0 (T) = \Delta^{-1} \sqrt{8 E_0 / \pi T}$ \cite{CurriePRB1980, *BogdanZPB1988}, which yields the blue solid line in Fig.~\ref{fig:fig4}. When $E_0 \to 0$, $\mathcal{D} \to \mathcal{D}_0 (L_\alpha / L)$; the algebraic decay is retained, provided that the temperature is well below the ordering temperature $T \ll T_c$ so that the conservation of the topological charge is maintained \cite{Note1}.

Thermal magnons, which have been disregarded in our treatment, can influence the diffusive motion of DWs and the associated spin transfer. Thermal magnons interact with DWs and can affect the diffusion coefficient $D$ at temperatures higher than the magnon's energy gap \cite{IvanovJPCM1993}. This could be captured by modifying the diffusion coefficient $D$, which enters in our main result, Eq.~(\ref{eq:main}). In addition, thermal magnons injected at the biased interface would exert a chirality-independent drag force on DWs within the spin-diffusion length \cite{BergerJAP1985, HatamiPRL2007, SaslowPRB2007, HalsSSC2010, HinzkePRL2011, YanPRL2011, ZhangPRL2012}. The associated change in the proposed DW spin transport is quadratic order in the bias,  and, therefore, the algebraic superfluid-like behavior of spin transport is not modified at the linear order in the bias.

There exist other materials with exotic spin phases supporting localized excitations with conserved ``charges," e.g., monopoles in spin ices with magnetic charges \cite{CastelnovoNature2008, *TchernyshyovNature2008, *GiblinNP2011}. These deconfined monopoles diffuse by Brownian motion as experimentally demonstrated \cite{BovoNC2013}, which leads us to envision that spin transport decaying algebraically may be implemented in a broader class of materials.

\begin{acknowledgments}
We are grateful to Scott Bender for helpful comments on the manuscript and to Rahul Roy for insightful discussions. This work was supported by the Army Research Office under Contract No. 911NF-14-1-0016 and in part by the U.S. Department of Energy, Office of Basic Energy Sciences under Award No. DE-SC0012190 and FAME (an SRC STARnet center sponsored by MARCO and DARPA).
\end{acknowledgments}

\bibliographystyle{/Users/evol/Dropbox/School/Research/apsrev4-1-nourl}
\bibliography{/Users/evol/Dropbox/School/Research/master}

\end{document}